\documentclass[prd,amsmath,amssymb,superscriptaddress,preprintnumbers,twocolumn,nofootinbib,10pt]{revtex4-1}
\usepackage{graphicx}
\usepackage{dcolumn}
\usepackage{bm}
\usepackage{amssymb}
\usepackage{latexsym}
\usepackage{booktabs}
\usepackage{amsmath}
\usepackage{multirow}
\usepackage{url}
\usepackage{footnote}
\usepackage{float}
\usepackage{threeparttable}
\usepackage[colorlinks=true, linkcolor=blue, citecolor=blue]{hyperref}
\usepackage[bottom]{footmisc}

\usepackage[normalem]{ulem}
\usepackage{color}
\usepackage{array}
\usepackage{enumerate}
\usepackage{adjustbox}

\usepackage{makecell}
\usepackage{diagbox}
\usepackage{epstopdf}
\usepackage{epsfig}
\usepackage{longtable}
\usepackage{supertabular}
\usepackage{algorithm}
\usepackage{pifont}
\usepackage{algorithmic}
\usepackage{changepage}
\usepackage{setspace}

\begin{document}

\title {Hints of noncold dark matter? Observational constraints on barotropic dark matter with a constant equation of state parameter}

\author{Yan-Hong Yao}\thanks{Corresponding author}
\email{yaoyanhong@nbu.edu.cn}
\affiliation{Institute of Fundamental Physics and Quantum Technology, Department of Physics, School of Physical Science and Technology, Ningbo University, Ningbo, Zhejiang 315211, China} 
\affiliation{School of Physics and Astronomy, Sun Yat-sen University, 2 Daxue Road, Tangjia, Zhuhai, People's Republic of China}
\author{Jian-Qi Liu}
\affiliation{School of Physics and Astronomy, Sun Yat-sen University, 2 Daxue Road, Tangjia, Zhuhai, People's Republic of China}
\begin{abstract}
This study investigates the potential of a cosmological model termed $\Lambda w$DM, in which a cosmological constant play the role of dark energy and dark matter is barotropic and has a constant equation of state parameter ($w_{\rm dm}$), to address the $S_8$ tension between early- and late- universe observations. By incorporating the latest cosmological datasets---including Planck Cosmic Microwave Background (CMB), Baryon Acoustic Oscillation (BAO), Ia supernovae (SNe Ia), Redshift Space Distortions (RSD), and weak lensing (WL)---we constrain the $\Lambda w$DM compared to 
$\Lambda$CDM. Our analysis reveals a marginal preference for a non-zero 
$w_{\rm dm}=2.7^{+2.0}_{-1.9}\times10^{-7}$(
at 95\% confidence level) when combining CMB, SDSS BAO, SNe Ia, RSD, and WL data, and a marginal preference for a non-zero 
$w_{\rm dm}=2.29^{+1.9}_{-2.0}\times10^{-7}$(
at 95\% confidence level) when combining CMB, DESI Y1 BAO, SNe Ia, RSD, and WL data. In addition, we find that, compared to $\Lambda$CDM, $\Lambda w$DM can alleviate the $S_8$ tension from 
$>3\sigma$ to $<1\sigma$. Furthermore, we find that, for CMB+SDSS+PP+RSD+WL
datasets, the  $\Lambda w$DM model is close to being positively preferred over the $\Lambda$CDM model.
\textbf{}
\end{abstract}

\maketitle

\section{Introduction}
\label{intro}
Dark matter (DM), an mysterious component of the universe that does not interact with photons, is known to make up approximately one-fourth of the universe's total energy content today. As the true nature of DM is still unknown, the prevailing DM model is largely phenomenological. In this model, DM is treated as a non-interacting perfect fluid with a zero equation of state (EoS) parameter, zero sound speed, and zero viscosity, commonly referred to as cold dark matter (CDM). CDM and the cosmological constant are two foundational components of the standard cosmological model, known as the $\Lambda$CDM model, which fits a wide range of cosmological observations across various scales~\citep{riess1998observational,perlmutter1999measurements,dunkley2011atacama,hinshaw2013nine,story2015measurement,alam2017clustering,troxel2018dark,aghanim2020planck-1}. Despite the success of this paradigm, it faces several small-scale challenges, such as the missing satellite problem~\citep{klypin1999missing,moore1999dark}, the too-big-to-fail problem~\citep{boylan2012milky}, and the core-cusp issue~\citep{moore1999cold,springel2008aquarius}. These persistent small-scale discrepancies have led to the development of alternative DM candidates beyond CDM, including warm DM~\citep{blumenthal1982galaxy,bode2001halo}, fuzzy DM~\citep{hu2000fuzzy,marsh2014model}, interacting DM~\citep{spergel2000observational}, and decaying DM~\citep{wang2014cosmological}, all of which aim to suppress the formation of low-mass structures while remaining consistent with large-scale observations.

Most of DM candidates can be described using the generalized dark matter (GDM) framework by assuming different parameterizations of the EoS parameter, sound speed, and viscosity. The GDM framework was first proposed in~\cite{hu1998structure}, and has since been followed by many other researchers~\citep{mueller2005cosmological, kumar2014observational, kopp2018dark, kumar2019testing, ilic2021dark, pan2023iwdm, yao2024observational, liu2025observational,li2025non}. In this work, we do not focus on several small-scale challenges mentioned earlier, instead, we aim to address the tension in $S_8=\sigma_8\sqrt{\Omega_{\rm m}/0.3}$ (where $\Omega_{\rm m}$ is the matter density parameter and $\sigma_8$ is the matter fluctuation amplitude on scales of $ 8h^{-1}{\rm Mpc}$) with the help of the $\Lambda$GDM (GDM and the cosmological constant play the role of DM and dark energy (DE), respectively) framework. This tension exists between high-redshift data, such as the Cosmic Microwave Background (CMB), and low-redshift data, including weak lensing (WL) and Large-Scale Structure (LSS)~\citep{macaulay2013lower,joudaki2016cfhtlens,bull2016beyond,joudaki2017kids,nesseris2017tension,kazantzidis2018evolution,asgari2020kids+,hildebrandt2020kids+,skara2020tension,abbott2020dark,joudaki2020kids+,heymans2021kids,asgari2021kids,loureiro2021kids,abbott2022dark,amon2022dark,secco2022dark,philcox2022boss}.
In order to achieve this goal in the simplest way, we first assume that the new DM candidate is non-viscous; secondly, we assume that the EoS parameter $w_{\rm dm}$ of the new DM candidate is constant; thirdly, we assume that the new DM candidate is barotropic, meaning its sound speed is equal to its adiabatic sound speed, since $w_{\rm dm}$ is constant, we further obtain that the square of its sound speed is equal to $w_{\rm dm}$. This new DM candidate is denoted as $w$DM. We believe that the new model can alleviate the $S_8$ tension because a small positive value of the sound speed of DM can slightly increase the Jeans wavelength, thereby appropriately reducing the matter power spectrum at small scales. Compared to the $\Lambda$CDM model, the new model introduces only one additional free parameter, namely, the $w_{\rm dm}$ parameter, and we refer to this new model as $\Lambda w$DM. In fact, although we refer to this as a new model, it has been proposed and discussed in Ref.~\cite{mueller2005cosmological} for about two decades. By utilizing the latest CMB, Type Ia supernovae, and large-scale structure data at the time, the author of Ref.~\cite{mueller2005cosmological} obtained $-11.9<10^7w_{\rm dm}<4.62$ at the 2$\sigma$ confidence level, which is consistent with the $\Lambda$CDM model. Given that the quantity and precision of current cosmological data far exceed those of the past, it is necessary for us to re-examine the $\Lambda$CDM model using the latest data in conjunction with this model.

This paper is organized as follows. Section~\ref{sec:1} outlines the key equations of the $\Lambda w$DM model. In section~\ref{sec:2}, we present the observational datasets and the statistical methodology. In section~\ref{sec:3}, we report the results and its implications regarding the $\Lambda w$DM model. The last section concludes.
\section{Review of the $\Lambda w$DM Model}
\label{sec:1}
At the scale of background level, the $\Lambda w$DM model is discussed in the context of a spatially flat, homogeneous, and isotropic spacetime, which is described by the spatially flat Friedmann-Robertson-Walker (FRW) metric. The evolution of this spacetime is assumed to be governed by general relativity and is therefore influenced by the components of the universe, which include the cosmological constant, $w$DM, baryons, and radiation. Any non-gravitational interactions among these components are neglected. Consequently, we can express the dimensionless Hubble parameter as
\begin{equation}\label{Eq:H}
E^2=\frac{H^{2}}{H_{0}^{2}} = \Omega_{\mathrm{r0}}(\frac{a_0}{a})^{4} + \Omega_{\mathrm{dm0}}(\frac{a_0}{a})^{3(1+w_{\rm dm})} + \Omega_{\mathrm{b0}}(\frac{a_0}{a})^{3} + \Omega_{\Lambda},
\end{equation}
where $H$ is the Hubble parameter, $a$ is the scale factor, the subscript 0 denotes the present moment. $\Omega_{\rm r0}$, $\Omega_{\rm dm0}$, $\Omega_{\rm b0}$, and $\Omega_{\Lambda}$ are the density parameters for radiation, $w$DM, baryons, and the cosmological constant, respectively. And these density parameters satisfy the relation $\Omega_{\Lambda}=1-\Omega_{\rm r0}-\Omega_{\rm dm0}-\Omega_{\rm b0}$.

At the scales of linear level, by adopting the conformal Newtonian gauge, the linear perturbation of the FRW metric is expressed in the following form
\begin{equation}
    ds^2=a^2(\tau)[-(1+2\psi)d\tau^2+(1-2\phi)d\vec{r}^2],
\end{equation}
where $\psi$ and $\phi$ represent the metric potentials, $\tau$ is the conformal time, while $\vec{r}$ denotes the three spatial coordinates. By considering the linear perturbation of the conserved stress-energy momentum tensor, we can derive the following continuity and Euler equations (in Fourier space) for $w$DM~\citep{kumar2019testing}
\begin{equation}\label{}
\delta_{\rm dm}^{\prime}= - (1+w_{\rm dm}) \left(\theta_{\rm dm}-3\phi^{\prime}\right)- 3 \mathcal{H} \left(\frac{\delta p_{\rm dm}}{\delta\rho_{\rm dm}} - w_{\rm dm}
	  \right)\delta_{\rm dm},
\end{equation}
\begin{equation}\label{}
  \theta_{\rm dm}^{\prime}=-\mathcal{H}
(1-3c_{\rm ad,dm}^2)\theta_{\rm dm}  + \frac{\delta p_{\rm dm}/\delta\rho_{\rm dm}}{1+w_{\rm dm}}k^2\delta_{\rm dm} + k^2\psi.
\end{equation}
Here, a prime stands for the conformal time derivative. The parameter $\mathcal{H}$ represents the conformal Hubble parameter, and k is the magnitude of the wavevector $\vec{k}$.
The terms $\delta_{\rm dm}$ and $\theta_{\rm dm}$ refer to the relative density and velocity divergence perturbations of $w$DM, respectively. The term $c_{\rm ad,dm}^2$ is the square of adiabatic sound speed of $w$DM, it is given by
\begin{equation}\label{}
  c_{\rm ad,dm}^2=\frac{p_{\rm dm}^{\prime}}{\rho_{dm}^{\prime}}=w_{\rm dm}-\frac{w_{\rm dm}^{\prime}}{3\mathcal{H}\left(1+w_{\rm dm}\right)},
\end{equation}
furthermore, $\frac{\delta p_{\rm dm}}{\delta\rho_{\rm dm}}$ is the square of sound speed of $w$DM in the Newtonian gauge, it can be expressed as
\begin{equation}
\label{deltaP}
\frac{\delta p_{\rm dm}}{\delta\rho_{\rm dm}}=c_{\rm s,dm}^2+3\mathcal{H}\left(1+w_{\rm dm}\right)\left(c_{\rm s,dm}^2-c_{\rm ad,dm}^2\right)\frac{\theta_{\rm dm}}{\delta_{\rm dm}k^2},
\end{equation}
here, $c_{\rm s,dm}^2=c_{\rm ad,dm}^2+c_{\rm nad,dm}^2$ is the square of sound speed of $w$DM in the rest frame. The term $c_{\rm nad,dm}^2$ is the square of non-adiabatic sound speed of $w$DM, which describes its micro-scale properties and needs to be provided independently. In this work, we consider $c_{\rm nad,dm}^2=0$. Therefore, we have $c_{\rm s,dm}^2=c_{\rm ad,dm}^2=w_{\rm dm}$. As a result, the continuity and Euler equations for $w$DM can be rewritten as follows
\begin{equation}\label{}
\delta_{\rm dm}^{\prime}= - (1+w_{\rm dm}) \left(\theta_{\rm dm}-3\phi^{\prime}\right),
\end{equation}

\begin{equation}\label{}
  \theta_{\rm dm}^{\prime}=-\mathcal{H}
(1-3w_{\rm dm})\theta_{\rm dm}  + \frac{w_{\rm dm}}{1+w_{\rm dm}}k^2\delta_{\rm dm} + k^2\psi.
\end{equation}

Having presented the equations above, the background and linear perturbation dynamics of the $\Lambda w$DM are clearly understood. We follow the conventions set by the Planck collaboration and model free-streaming neutrinos as consisting of two massless species and one massive species with a mass of $M_v=0.06$~eV. Therefore, the full baseline
parameters set of $\Lambda w$DM is given by
\begin{equation}\label{}
	\mathcal{P}=\{\omega_b, \omega_{\rm dm}, \theta_s, A_s, n_s, \tau_{\rm reio}, w_{\rm dm}\}.
\end{equation}

\section{Data sets and methodology}
\label{sec:2}
To extract the free parameters of the $\Lambda w$DM model, we use the recent observational datasets described below.

\textbf{Cosmic Microwave Background (CMB)}: we utilize the baseline of Planck 2018~\citep{aghanim2020planck-1,aghanim2020planck-2}, more specifically, we use the CMB temperature
and polarization angular power spectra plikTTTEEE+lowE+lowl. Additionally, we incorporate the Planck 2018 CMB lensing reconstruction likelihood~\citep{aghanim2020planck-3} into our analysis.

\textbf{Baryon Acoustic Oscillation (BAO)}: in our analysis of BAO datasets, we will focus on two distinct samples: one derived from the Six-degree Field Galaxy Survey (6dFGS) and the Sloan Digital Sky Survey (SDSS), and the other from recent observations by the Dark Energy Spectroscopy Instrument (DESI).
\begin{center}
1.BAO Measurements from 6dFGS and SDSS 
\end{center}
This sample encompasses the following BAO measurements: (1) the measurement obtained from 6dFGS at an effective redshift of $z_{\rm eff}$= 0.106~\cite{Beutler2011The}; (2) the BAO-only portion of the eBOSS DR16 compilation~\cite{alam2021completed}, which incorporates data from SDSS DR7 MGS~\cite{ross2015clustering}, BOSS DR12~\cite{alam2017clustering}, as well as eBOSS DR16 Luminous Red Galaxy (LRG) samples~\cite{bautista2021completed,gil2020completed}, eBOSS DR16 Quasar (QSO) samples~\cite{hou2021completed}, eBOSS DR16 Emission Line Galaxies (ELG) samples~\cite{de2021completed}, and eBOSS DR16 Ly$\alpha$ forest samples~\cite{des2020completed}. We refer to this sample as SDSS.
\begin{center}
2. BAO Measurements from DESI 
\end{center}
This sample includes the BAO measurements obtained from the first year of DESI observations, as detailed in Ref.~\cite{adame2024desi1}. It consists of the Bright Galaxy Sample (BGS), LRG, a combination of LRG and ELG, ELG, QSO, and Ly$\alpha$ forest~\cite{adame2024desi1,adame2024desi2,adame2024desi3}. We refer to this sample as DESI.

\textbf{Pantheon Plus (PP)}: We use the distance modulus measurements of Type Ia supernovae (SNe Ia) from the Pantheon+ sample, as detailed in~\citep{scolnic2022pantheon+}. This dataset comprises 1701 light curves corresponding to 1550 distinct SNe Ia events, spanning the redshift range of $z\in[0.001, 2.26]$. This dataset is commonly referred to as Pantheon Plus (PP).

\textbf{Redshift Space Distortions (RSD)}: RSD is caused by the peculiar velocities of objects along the line of sight, resulting in a mapping from real space to redshift space. This effect introduces anisotropies in the clustering patterns of objects and is affected by the growth of cosmic structures, making RSD a sensitive probe for the combination $f\sigma_8$. In the $\Lambda w$DM model, the quantity $f$ is scale-dependent and can be defined as:
\begin{equation}\label{}
  f(k,a)=\frac{d\ln\delta(k,a)}{d\ln a}, \hspace{1cm} \delta(k,a)=\sqrt{\frac{P(k,a)}{P(k,a_0)}},
\end{equation}
where $P(k,a)$ represents the matter power spectrum, defined by:
\begin{equation}\label{}
  \langle\widetilde{\delta}_m(\textbf{k},t)\widetilde{\delta}_m^\ast(\textbf{k}^{\prime},t)\rangle=(2\pi)^3P(\textbf{k},a(t))\delta^3(\textbf{k}-\textbf{k}^{\prime}),
\end{equation}
here $\widetilde{\delta}_m(\textbf{k},t)$ is the Fourier transform of $\delta_m=\frac{\delta\rho_{\rm dm}+\delta\rho_b+\delta\rho_{\rm \overline{\nu}}}{\bar{\rho}_{\rm dm}+\bar{\rho}_b+\bar{\rho}_{\rm \overline{\nu}}}$, with the subscript $\overline{\nu}$ indicating massive neutrinos. The quantity $\sigma_8$ is given by:
\begin{equation}\label{}
  \sigma_8(a)=\sqrt{\int_0^{+\infty}dk\frac{k^2P(k,a)W_R^2(k)}{2\pi^2}}
\end{equation}
where $W_R(k) = 3[\sin(kR)/kR-\cos(kR)]/(kR)^2$ is the Fourier transform of the top-hat window function, and $R$ denotes the scale for calculating the root-mean-square (RMS) normalization of matter fluctuations.

In this study, we will utilize the RSD measurements of $f\sigma_8$ presented in Table I of Ref.~\cite{sagredo2018internal}. This dataset includes 22 values of $f\sigma_8$ spanning the redshift range $0.02<z<1.944$, sourced from a variety of surveys: 2dFGRS~\cite{song2009reconstructing}, 2MASS~\cite{davis2011local}, SDSS-II LRGs~\cite{samushia2012interpreting}, First Amendment SNeIa+IRAS~\cite{turnbull2012cosmic,hudson2012growth}, WiggleZ~\cite{blake2012wigglez}, GAMA~\cite{blake2013galaxy}, BOSS DR11 LOWZ~\cite{sanchez2014clustering}, BOSS DR12 CMASS~\cite{chuang2016clustering}, SDSS DR7 MGS~\cite{howlett2015clustering}, SDSS DR7~\cite{feix2015growth}, FastSound~\cite{okumura2016subaru}, Supercal SNeIa+6dFGS~\cite{huterer2017testing}, VIPERS PDR-2~\cite{pezzotta2017vimos}, and eBOSS DR14 quasars~\cite{zhao2019clustering}. These measurements are collectively referred to as the "Gold 2018" sample in the literature. For this article, we will use the publicly available Gold 2018 Montepython likelihood~\cite{arjona2020cosmological} for the $f\sigma_8$ dataset. In this likelihood, the wavenumber $k$ in $f$ is fixed at 0.1 Mpc, consistent with the effective wavenumber of the RSD measurements utilized.

\textbf{Weak Lensing (WL)}: in addition to the datasets mentioned above, we incorporate a prior on $S_8$, i.e.~$S_8=0.759^{+0.024}_{-0.021}$~\cite{asgari2021kids}, which is based on the measurements from KiDS1000.
(For the $\Lambda w$DM model, utilizing the full WL likelihood necessitates a thorough consideration of nonlinear effects. Due to the unavailability of these tools, we confine our analysis to the linear power spectrum and assume that including the $S_8$ prior adequately represents the constraints imposed by the KiDS1000 likelihood on the $\Lambda w$DM model).

To constrain the $\Lambda w$DM model, we run a Markov Chain Monte Carlo (MCMC) using the public MontePython-v3 code~\cite{audren2013conservative,brinckmann2019montepython}, which is interfaced with a modified version of the CLASS code~\cite{lesgourgues2011cosmic,blas2011cosmic}. Our analysis employs the Metropolis-Hastings algorithm, and we consider the chains to have converged when the Gelman-Rubin criterion~\cite{gelman1992inference} $R-1<0.01$ is satisfied. In Tab.~\ref{tab:prior} we display the flat priors on the free parameters of $\Lambda w$DM. In particular, since the condition $c_{\rm s,dm}^2=w_{\rm dm}<0$ will lead to an imaginary value of sound speed of DM, we impose a lower bound of 0 for the flat prior on the parameter $w_{\rm dm}$. Finally, we have used the GetDist Python package~\cite{Lewis:2019xzd} to analyze the samples.

\begin{table}[!t]
	\centering
	\renewcommand\arraystretch{1.2}
	\setlength{\tabcolsep}{3mm}{
		\begin{ruledtabular}
			\begin{tabular}{c c}
				Parameters               & Prior       \\
				\hline
				$100\omega{}_{b}$        & [0.8,2.4]   \\
				$\omega{}_{\mathrm{dm}}$ & [0.01,0.99] \\
				$100\theta{}_{s}$        & [0.5,2.0]   \\
				$\ln[10^{10}A_{s}]$      & [2.7,4.0]   \\
				$n_{s}$                  & [0.9,1.1]   \\
				$\tau{}_{\mathrm{reio}}$ & [0.01,0.8]  \\
				$10^{6}w_{\mathrm{dm}}$  & [0,100]     \\
			\end{tabular}
	\end{ruledtabular}}
	\caption{Uniform priors on the free parameters of the $\Lambda w$DM model.}
	\label{tab:prior}
\end{table}

\section{Results and discussion}
\label{sec:3}
\begin{figure*}
	\centering
	\includegraphics[scale=0.5]{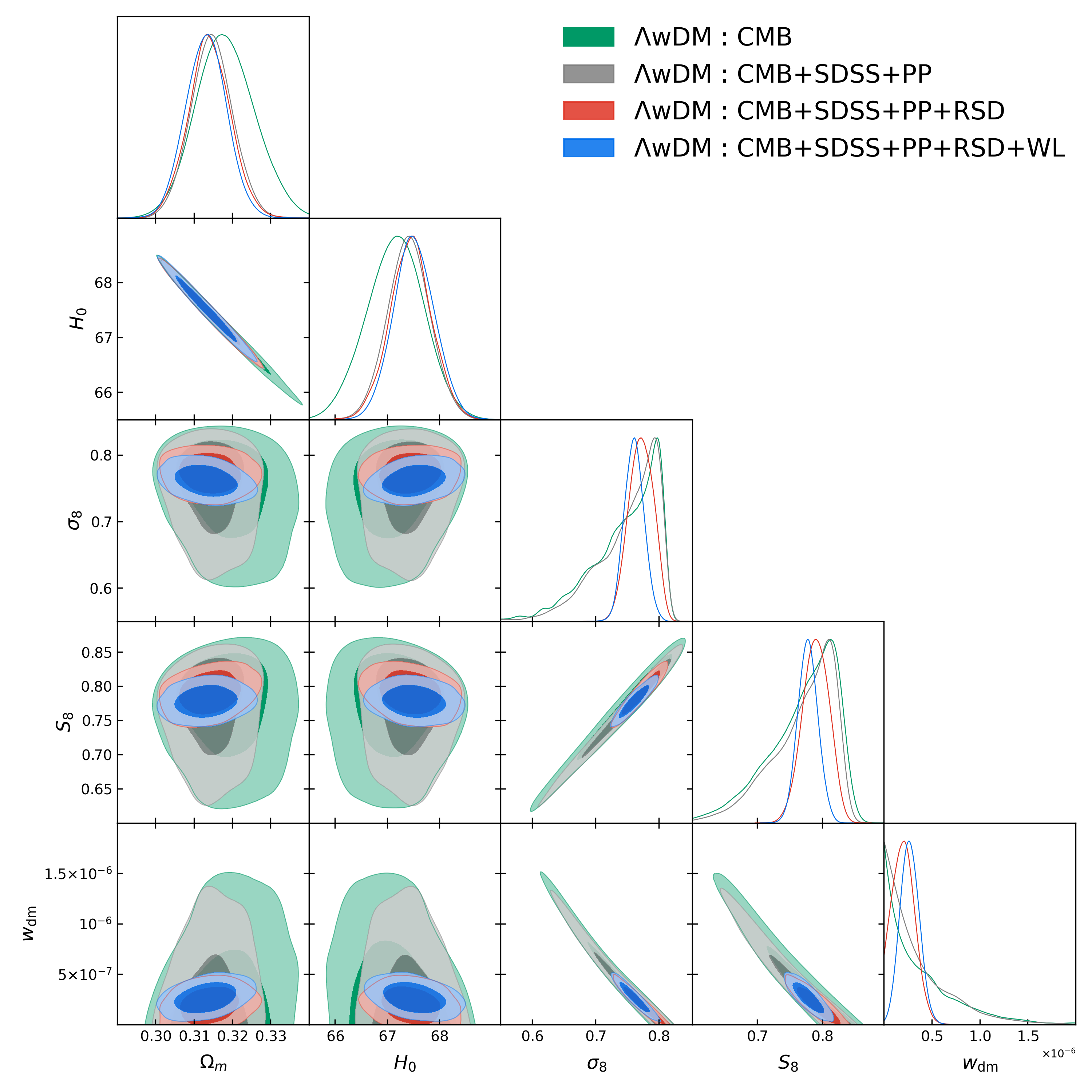}
	\caption{One dimensional posterior distributions and two dimensional joint contours at 68\% and 95\% CL for the most relevant parameters of the $\Lambda w$DM model using CMB, CMB+SDSS+PP, CMB+SDSS+PP+RSD, and CMB+SDSS+PP+RSD+WL datasets. }
\label{fig:1}
\end{figure*}

\begin{figure*}
	\centering
	\includegraphics[scale=0.4]{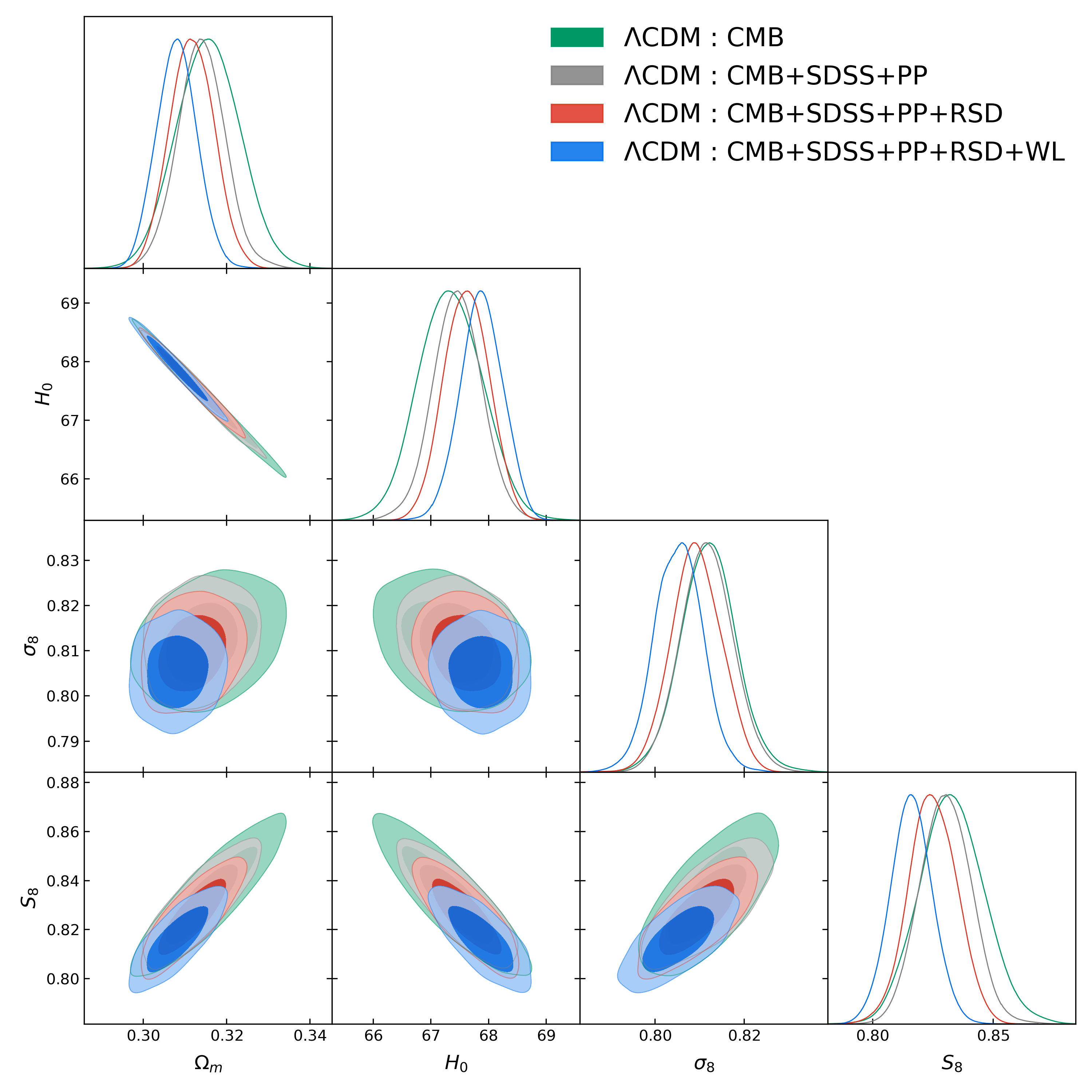}
	\caption{One dimensional posterior distributions and two dimensional joint contours at 68\% and 95\% CL for the most relevant parameters of the $\Lambda$CDM model using CMB, CMB+SDSS+PP, CMB+SDSS+PP+RSD, and CMB+SDSS+PP+RSD+WL datasets.}
\label{fig:2}
\end{figure*}

In Tab.~\ref{tab:LwDM_SDSS} and Fig.~\ref{fig:1}, we present the constraints on the $\Lambda w$DM model for CMB, CMB+SDSS, CMB+SDSS+PP,  CMB+SDSS+PP+RSD, and CMB+SDSS+PP+RSD+WL datasets. Additionally, we provide the fitting results of the $\Lambda$CDM model for the same datasets in Tab.~\ref{tab:LCDM_SDSS}  and Fig.~\ref{fig:2} for comparison. 

We begin by analyzing the fitting results of the $\Lambda w$DM model using only CMB data. Next, we investigate the effects of including additional probes by gradually incorporating them into the CMB analysis. When using solely CMB data, we find no indication of a non-zero DM parameter $w_{\rm dm}$ at the 68\% confidence level. Nevertheless, the differences in some model parameters between the $\Lambda w$DM model and the $\Lambda$CDM model are shown, attributable to the small yet significant positive mean value of $w_{\rm dm}$. More specifically, the small but non-zero positive mean value of the parameter $w_{\rm dm}$ causes a decrease in the parameters $\sigma_8$ and $S_8$, shifting their values from $\sigma_8=0.8121\pm0.0062$ (at the 68\% confidence level) and $S_8=0.833\pm0.013$ (at the 68\% confidence level) in the $\Lambda$CDM model to $\sigma_8=0.747^{+0.065}_{-0.023}$ (at the 68\% confidence level) and $S_8=0.770^{+0.066}_{-0.027}$ (at the 68\% confidence level) in the $\Lambda w$DM model. The reason that such a small positive value of $w_{\rm dm}$ can lead to moderate changes in the values of $\sigma_8$($S_8$) is that a small positive value of sound speed of DM can moderately increase the Jeans wavelength, thereby appropriately reducing the matter power spectrum at small scales, this causal relationship is demonstrated by the negative correlation between between $\sigma_8$($S_8$) and $w_{\rm dm}$, as is shown in Fig.~\ref{fig:1}. 

With the addition of the SDSS+PP dataset to the CMB, we observe little change in the fitting results compared to the CMB analysis results. This is because, firstly, BAO and SN Ia data are background data, so they do not impose constraints on the sound speed of DM. Secondly, the error bars for both types of data are relatively large at present, so their constraints on the EoS parameter of DM, as background physical quantity, are not particularly tight.

When RSD are added to CMB+SDSS+PP, we observe that there is an indication of a non-zero DM parameter $w_{\rm dm}$ at the 68\% confidence level. However, 1$\sigma$ level is not statistically significant. Furthermore, RSD dataset slightly improves the values of parameters $\sigma_8$ and $S_8$ compared to that regarding the CMB+SDSS+PP analysis.

For the CMB+SDSS+PP+RSD+WL dataset, we finally find a statistically significant signal for a positive DM parameter $w_{\rm dm}=2.70^{+2.0}_{-1.9}\times10^{-7}$ (at the 95\% confidence level). However, we still need to be cautious about this result, because, since we currently do not have a halo model for $\Lambda w$DM, we can only use the model-dependent $S_8$ prior to replace the complete WL likelihood to constrain $\Lambda w$DM. This may introduce some bias into the fitting results. Furthermore, Eq.~\ref{Eq:H} can be rewritten as \begin{equation}
	E^2= \Omega_{\mathrm{r0}}(\frac{a_0}{a})^{4} + \Omega_{\mathrm{m0}}(\frac{a_0}{a})^{3} + \Omega_{\Lambda}+\Omega_{\mathrm{dm0}}(\frac{a_0}{a})^{3}((\frac{a_0}{a})^{3w_{\rm dm}}-1).
\end{equation}
Therefore, the background evolution of the $\Lambda w$DM model can be equivalently considered as being determined by radiation, baryons, CDM, and a dynamical DE whose energy density is parameterized as $\Omega_{\Lambda}+\Omega_{\mathrm{dm0}}(\frac{a_0}{a})^{3}((\frac{a_0}{a})^{3w_{\rm dm}}-1)$. As long as $w_{\rm dm}>0$, this dynamical DE behaves like a cosmological constant at low redshifts and like a quintessence at high redshifts, which is in tension with recently released DESI data since the latter support phantom behavior for DE at high redshifts. Therefore, if we replace the SDSS data with the DESI data, the new data could prevent the EoS parameter of the dynamical DE from deviating significantly towards values greater than $-1$, and may cause the parameter $w_{\rm dm}$ to no longer be greater than 0 at the 2$\sigma$ confidence level.
 
\begin{figure*}
	\centering
	\includegraphics[scale=0.5]{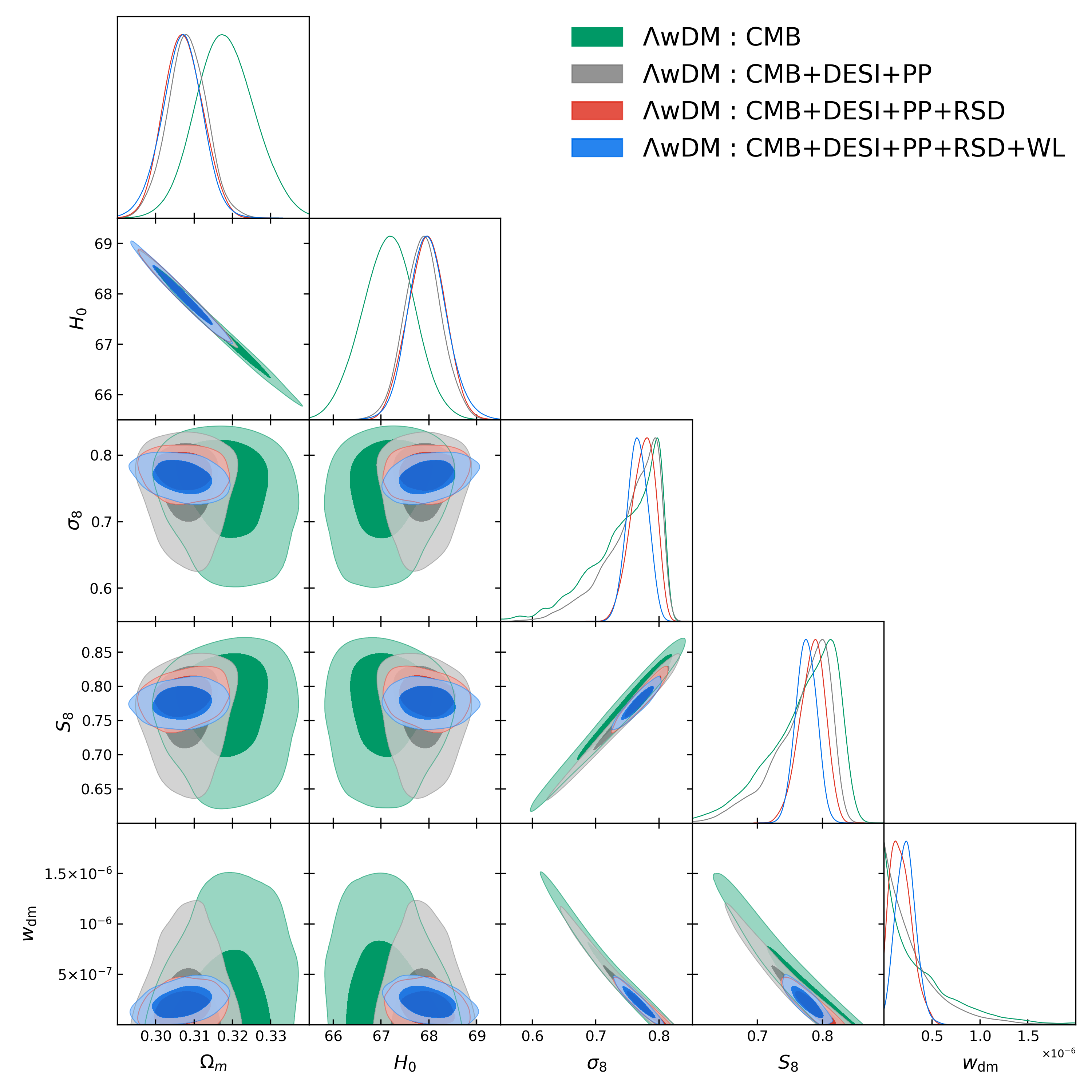}
	\caption{One dimensional posterior distributions and two dimensional joint contours at 68\% and 95\% CL for the most relevant parameters of the $\Lambda w$DM model using CMB, CMB+DESI+PP, CMB+DESI+PP+RSD, and CMB+DESI+PP+RSD+WL datasets. }
\label{fig:3}
\end{figure*}

\begin{figure*}
	\centering
	\includegraphics[scale=0.4]{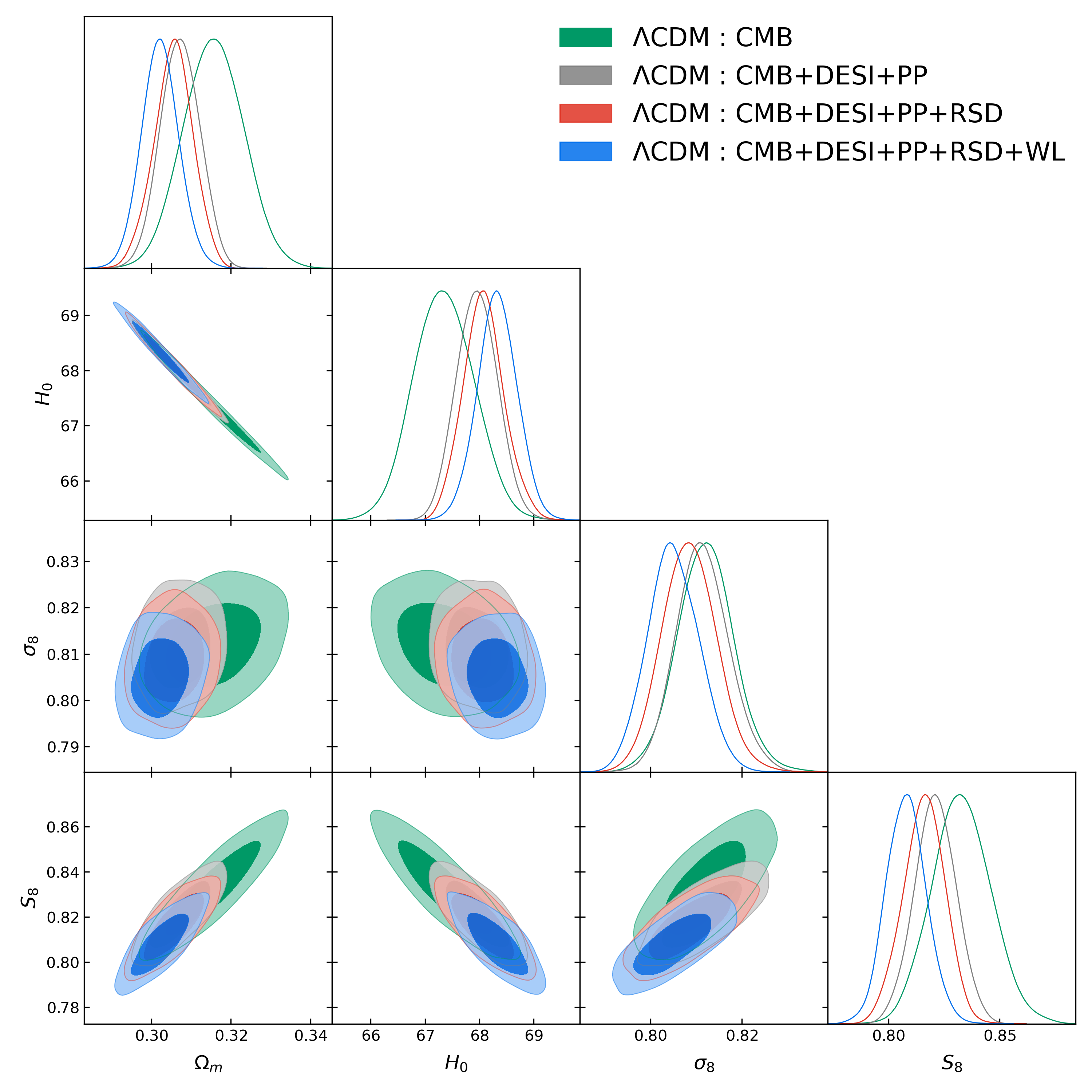}
	\caption{One dimensional posterior distributions and two dimensional joint contours at 68\% and 95\% CL for the most relevant parameters of the $\Lambda$CDM model using CMB, CMB+DESI+PP, CMB+DESI+PP+RSD, and CMB+DESI+PP+RSD+WL datasets. }
\label{fig:4}
\end{figure*}

To test this idea, we replace the SDSS data with DESI data in the previous three datasets that include BAO data and use them to constrain the $\Lambda w$DM and $\Lambda$CDM models. We present the corresponding fitting results in Tab.~\ref{tab:LwDM_SDSS}, Fig.~\ref{fig:3} and Tab.~\ref{tab:LCDM_SDSS},  Fig.~\ref{fig:4}. We find that the results show no significant changes compared to the previous cases. In particular, although the mean values of $w_{\rm dm}$ derived from the three new datasets decrease slightly, there is still a statistically significant signal for a positive DM parameter, $w_{\rm dm}=2.29^{+1.9}_{-2.0}\times10^{-7}$ (at the 95\% confidence level), for the CMB+DESI+PP+RSD+WL datasets. 

\begin{figure*}
	\centering
	\includegraphics[scale=0.4]{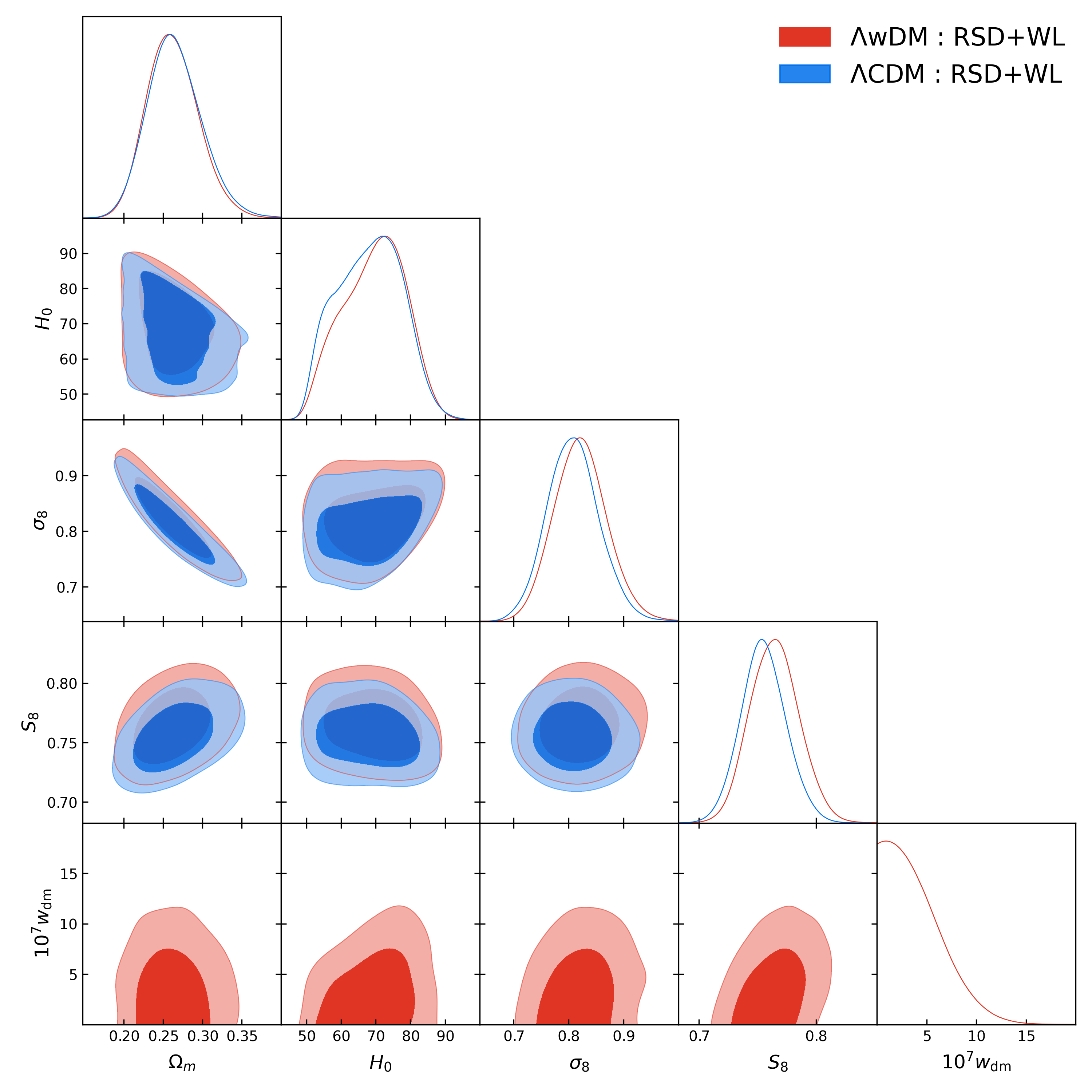}
	\caption{One dimensional posterior distributions and two dimensional joint contours at 68\% and 95\% CL for the most relevant parameters of the $\Lambda w$DM  using RSD+WL datasets, compared with the results from 
    $\Lambda$CDM model. }
	\label{fig:5}
\end{figure*}

Before we discuss the $S_8$ problem, we present the fitting results of the $\Lambda w$DM and $\Lambda$CDM models for the RSD+WL datasets as well as a Gaussian prior on $\omega_b$ from Big Bang Nucleosynthesis (BBN): 100$\omega_b=2.233\pm0.036$~\cite{mossa2020baryon} in Tab.~\ref{tab:RSD+WL}  and Fig.~\ref{fig:5}. We refer to this data combination as RSD+WL.   Now, we assess the $\Lambda w$DM model's ability to relieve the $S_8$ tension by using the following two quantities
\begin{equation*}
	T_1=\frac{x_{\rm CMB+SDSS+PP}-x_{\rm RSD+WL}}{\sqrt{\sigma_{\rm CMB+SDSS+PP}^2+\sigma_{\rm RSD+WL}^2}}, 
\end{equation*}
\begin{equation*}
	T_2=\frac{x_{\rm CMB+DESI+PP}-x_{\rm RSD+WL}}{\sqrt{\sigma_{\rm CMB+DESI+PP}^2+\sigma_{\rm RSD+WL}^2}}, 
\end{equation*}
where $x=S_8$. For the $\Lambda w$DM model, we have $T_1=0.16$ and $T_2=0.10$; for the $\Lambda$CDM model, we have $T_1=3.42$ and $T_2=3.12$, therefore, we can see that
the $\Lambda w$DM model can reduce the $S_8$ tension from beyond 3$\sigma$ to below 1$\sigma$.

Finally, we use the Akaike Information Criteria (AIC)~\citep{akaike1974new}, defined as
\begin{equation*}
{\rm AIC}=-2\ln \mathcal{L}_{\rm max}+2N=\chi^2_{\rm min}+2N,
\end{equation*}
where $\mathcal{L}_{\rm max}$ and $N$ denote the maximum likelihood and the total number of independent free parameters in the model, to compare the $\Lambda w$DM model with the $\Lambda$CDM model by computing the AIC
difference between two models, i.e., $ \triangle {\rm AIC}_{\Lambda w {\rm DM},\Lambda {\rm CDM}}={\rm AIC}_{\Lambda w {\rm DM}}-{\rm AIC}_{\Lambda {\rm CDM}} $. In alignment with ~\citep{schoneberg2022h0}, we demand that the preference for the $\Lambda w$DM model over the $\Lambda$CDM model is larger than a "weak preference" on Jeffrey's scale~\citep{jeffreys1961edition,nesseris2013jeffreys},
which leads to the criterion $ \triangle {\rm AIC}_{\Lambda w {\rm DM},\Lambda {\rm CDM}}\leq-6.91.$

Tab.~\ref{tab:AIC} summarizes the $ \triangle {\rm AIC}_{\Lambda w {\rm DM},\Lambda {\rm CDM}}$ values for all the data combinations. We observe that, for all the datasets, the difference between the models is not statistically significant, i.e., one of them isn't supported more than the other beyond a 'weak preference' on Jeffrey's scale. Nevertheless, for CMB+SDSS+PP+RSD+WL datasets, the $\Lambda w$DM model is close to being positively preferred over the $\Lambda$CDM model, this is shown by the
absolute value of $ \triangle {\rm AIC}_{\Lambda w {\rm DM},\Lambda {\rm CDM}}$ that larger than 6.

\begin{table*}[!t]
    \centering
    \renewcommand\arraystretch{1.5}
    \setlength{\tabcolsep}{1mm}{
        \begin{ruledtabular}
            \begin{tabular}{ccccc}
                Parameters                           & CMB                                              & CMB+SDSS+PP                                       & CMB+SDSS+PP+RSD                              & CMB+SDSS+PP+RSD+WL                           \\
                \hline
                {\boldmath$100\omega{}_{b}$}      & ${2.232\pm 0.015}^{+0.029}_{-0.028}$             & ${2.237\pm0.014}^{+0.027}_{-0.027}$              & ${2.237\pm0.013}^{+0.025}_{-0.026}$         & ${2.238\pm0.013}^{+0.027}_{-0.026}$         \\
                {\boldmath$\omega{}_{\mathrm{dm}}$} & ${0.1205\pm 0.0012}^{+0.0025}_{-0.0024}$         & ${0.11995\pm0.00090}^{+0.0018}_{-0.0018}$        & ${0.11985\pm0.00091}^{+0.0018}_{-0.0018}$   & ${0.11972\pm0.00087}^{+0.0017}_{-0.0017}$   \\
                {\boldmath$100\theta{}_{s}$}         & ${1.04187\pm 0.00028}^{+0.00053}_{-0.00056}$     & ${1.04190\pm0.00028}^{+0.00058}_{-0.00053}$      & ${1.04190\pm0.00027}^{+0.00053}_{-0.00055}$ & ${1.04192\pm0.00028}^{+0.00053}_{-0.00053}$ \\
                {\boldmath$\ln(10^{10}A_{s})$}          & ${3.048\pm 0.014}^{+0.029}_{-0.028}$             & ${3.051^{+0.013}_{-0.015}}^{+0.031}_{-0.028}$    & ${3.049\pm0.014}^{+0.028}_{-0.028}$         & ${3.050\pm0.014}^{+0.030}_{-0.027}$         \\
                {\boldmath$n_{s}$}                   & ${0.9637\pm 0.0040}^{+0.0079}_{-0.0078}$         & ${0.9651\pm0.0036}^{+0.0073}_{-0.0071}$          & ${0.9655\pm0.0037}^{+0.0073}_{-0.0070}$     & ${0.9655\pm0.0036}^{+0.0072}_{-0.0072}$     \\
                {\boldmath$\tau{}_{\mathrm{reio}}$}  & ${0.0553\pm 0.0074}^{+0.015}_{-0.015}$           & ${0.0571^{+0.0067}_{-0.0079}}^{+0.015}_{-0.014}$ & ${0.0564\pm0.0071}^{+0.014}_{-0.014}$       & ${0.0569\pm0.0072}^{+0.015}_{-0.013}$       \\
                {\boldmath$10^{7}w_{\mathrm{dm}}$}   & ${4.1^{+0.73}_{-4.1}}^{+8.1}_{-4.1}$                                   & ${3.64^{+0.54}_{-3.64}}^{+7.16}_{-3.64}$                                     & ${2.1^{+1.0}_{-1.3}}^{+1.94}_{-2.1}$                  & ${2.70\pm0.99}^{+2.0}_{-1.9}$               \\
                \hline
                {\boldmath$H_0$}                     & ${67.16\pm 0.55}^{+1.1}_{-1.1}$                  & ${67.41\pm0.41}^{+0.80}_{-0.79}$                 & ${67.45\pm0.40}^{+0.79}_{-0.81}$            & ${67.52\pm0.40}^{+0.79}_{-0.76}$            \\
                {\boldmath$\Omega{}_{m}$}            & ${0.3183^{+0.0072}_{-0.0081}}^{+0.015}_{-0.015}$ & ${0.3146\pm0.0055}^{+0.011}_{-0.011}$            & ${0.3141\pm0.0055}^{+0.011}_{-0.011}$       & ${0.3132\pm0.0053}^{+0.010}_{-0.010}$       \\
                {\boldmath$\sigma_8$}                & ${0.747^{+0.065}_{-0.023}}^{+0.072}_{-0.11}$     & ${0.752^{+0.059}_{-0.021}}^{+0.068}_{-0.099}$    & ${0.772\pm0.019}^{+0.036}_{-0.037}$         & ${0.761\pm0.016}^{+0.031}_{-0.029}$         \\
                {\boldmath$S_8$}                     & ${0.770^{+0.066}_{-0.027}}^{+0.081}_{-0.11}$     & ${0.770^{+0.061}_{-0.023}}^{+0.073}_{-0.10}$     & ${0.790\pm0.020}^{+0.039}_{-0.040}$         & ${0.778\pm0.015}^{+0.031}_{-0.030}$         \\         \hline 
                {\boldmath$\chi^2_{\rm min}$}           &
                $2782.18$  &  $ 4206.06 $  &  $4219.92$ 
                & $4220.58$
                 \\
            \end{tabular}
        \end{ruledtabular}}
    \caption{The mean values and 1, 2$\sigma$ of the $\Lambda w$DM model for CMB, CMB+SDSS+PP, CMB+SDSS+PP+RSD and CMB+SDSS+PP+RSD+WL datasets.}
    \label{tab:LwDM_SDSS}
\end{table*}

\begin{table*}[!t]
    \centering
    \renewcommand\arraystretch{1.5}
    \setlength{\tabcolsep}{1mm}{
        \begin{ruledtabular}
            \begin{tabular}{ccccc}
                Parameters                           & CMB                                           & CMB+SDSS+PP                                         & CMB+SDSS+PP+RSD                              & CMB+SDSS+PP+RSD+WL                             \\
                \hline
                {\boldmath$100\omega{}_{b}$}      & ${2.237^{+0.014}_{-0.016}}^{+0.030}_{-0.028}$ & ${2.239\pm0.014}^{+0.027}_{-0.028}$                & ${2.241\pm0.013}^{+0.025}_{-0.026}$         & ${2.245^{+0.013}_{-0.014}}^{+0.026}_{-0.024}$ \\
                {\boldmath$\omega{}_{\mathrm{cdm}}$} & ${0.1201\pm 0.0012}^{+0.0024}_{-0.0024}$      & ${0.11985\pm0.00092}^{+0.0019}_{-0.0019}$          & ${0.11949\pm0.00089}^{+0.0018}_{-0.0017}$   & ${0.11887\pm0.00080}^{+0.0016}_{-0.0015}$     \\
                {\boldmath$100\theta{}_{s}$}         & ${1.04189\pm 0.00030}^{+0.00059}_{-0.00059}$  & ${1.04189\pm0.00031}^{+0.00059}_{-0.00060}$        & ${1.04194\pm0.00029}^{+0.00057}_{-0.00059}$ & ${1.04194\pm0.00026}^{+0.00049}_{-0.00053}$   \\
                {\boldmath$\ln(10^{10}A_{s})$}          & ${3.046\pm 0.015}^{+0.030}_{-0.030}$          & ${3.047^{+0.013}_{-0.015}}^{+0.028}_{-0.027}$      & ${3.044\pm0.014}^{+0.028}_{-0.026}$         & ${3.039\pm0.014}^{+0.028}_{-0.027}$           \\
                {\boldmath$n_{s}$}                   & ${0.9650\pm 0.0041}^{+0.0079}_{-0.0082}$      & ${0.9654^{+0.0039}_{-0.0034}}^{+0.0072}_{-0.0085}$ & ${0.9662\pm0.0037}^{+0.0073}_{-0.0077}$     & ${0.9670\pm0.0036}^{+0.0071}_{-0.0070}$       \\
                {\boldmath$\tau{}_{\mathrm{reio}}$}  & ${0.0547\pm 0.0075}^{+0.015}_{-0.015}$        & ${0.0556\pm0.0072}^{+0.015}_{-0.014}$              & ${0.0545\pm0.0071}^{+0.015}_{-0.014}$       & ${0.0531\pm0.0071}^{+0.014}_{-0.014}$         \\
                \hline
                {\boldmath$H_0$}                     & ${67.35\pm 0.55}^{+1.1}_{-1.0}$               & ${67.46\pm0.42}^{+0.85}_{-0.85}$                   & ${67.62\pm0.39}^{+0.75}_{-0.78}$            & ${67.88\pm0.37}^{+0.70}_{-0.75}$              \\
                {\boldmath$\Omega{}_{m}$}            & ${0.3156\pm 0.0076}^{+0.015}_{-0.015}$        & ${0.3140\pm0.0057}^{+0.011}_{-0.011}$              & ${0.3118\pm0.0053}^{+0.010}_{-0.010}$       & ${0.3081\pm0.0048}^{+0.0094}_{-0.0094}$       \\
                {\boldmath$\sigma_8$}                & ${0.8121\pm 0.0062}^{+0.013}_{-0.012}$        & ${0.8117\pm0.0059}^{+0.012}_{-0.012}$              & ${0.8094\pm0.0056}^{+0.011}_{-0.011}$       & ${0.8054\pm0.0054}^{+0.011}_{-0.010}$         \\
                {\boldmath$S_8$}                     & ${0.833\pm 0.013}^{+0.027}_{-0.025}$          & ${0.830\pm0.011}^{+0.020}_{-0.020}$                & ${0.8251\pm0.0099}^{+0.019}_{-0.020}$       & ${0.8161\pm0.0086}^{+0.018}_{-0.017}$         \\ \hline 
                {\boldmath$\chi^2_{\rm min}$}           &
                $2781.58$  &  $ 4205.90 $  &  $4223.08$ 
                & $4228.98$
                \\
            \end{tabular}
        \end{ruledtabular}}
    \caption{The mean values and 1, 2$\sigma$ of the $\Lambda$CDM model for CMB, CMB+SDSS+PP, CMB+SDSS+PP+RSD and CMB+SDSS+PP+RSD+WL datasets.}
    \label{tab:LCDM_SDSS}
\end{table*}

\begin{table*}[!t]
    \centering
    \renewcommand\arraystretch{1.5}
    \setlength{\tabcolsep}{1mm}{
        \begin{ruledtabular}
            \begin{tabular}{cccc}
                Parameters                                                                      & CMB+DESI+PP                                      & CMB+DESI+PP+RSD                                  & CMB+DESI+PP+RSD+WL                                    \\
                \hline
                {\boldmath$100\omega{}_{b}$}                & ${2.244\pm0.014}^{+0.027}_{-0.026}$              & ${2.245\pm0.013}^{+0.027}_{-0.027}$              & ${2.246\pm0.014}^{+0.028}_{-0.027}$                   \\
                {\boldmath$\omega{}_{\mathrm{dm}}$}          & ${0.11893\pm0.00087}^{+0.0017}_{-0.0018}$        & ${0.11876\pm0.00084}^{+0.0017}_{-0.0016}$        & ${0.11870^{+0.00091}_{-0.00081}}^{+0.0017}_{-0.0018}$ \\
                {\boldmath$100\theta{}_{s}$}            & ${1.04203\pm0.00027}^{+0.00054}_{-0.00053}$      & ${1.04204\pm0.00028}^{+0.00055}_{-0.00057}$      & ${1.04202\pm0.00028}^{+0.00055}_{-0.00055}$           \\
                {\boldmath$\ln(10^{10}A_{s})$}                    & ${3.054^{+0.014}_{-0.016}}^{+0.031}_{-0.028}$    & ${3.054^{+0.014}_{-0.016}}^{+0.031}_{-0.028}$    & ${3.053\pm0.015}^{+0.030}_{-0.029}$                   \\
                {\boldmath$n_{s}$}                        & ${0.9679\pm0.0037}^{+0.0075}_{-0.0071}$          & ${0.9682\pm0.0036}^{+0.0070}_{-0.0074}$          & ${0.9684\pm0.0036}^{+0.0073}_{-0.0070}$               \\
                {\boldmath$\tau{}_{\mathrm{reio}}$}            & ${0.0600^{+0.0068}_{-0.0080}}^{+0.016}_{-0.015}$ & ${0.0600^{+0.0069}_{-0.0080}}^{+0.016}_{-0.014}$ & ${0.0598^{+0.0069}_{-0.0080}}^{+0.016}_{-0.014}$      \\
                {\boldmath$10^{7}w_{\mathrm{dm}}$}                                    & ${3.13^{+0.42}_{-3.13}}^{+6.17}_{-3.13}$                                     & ${1.84^{+0.74}_{-1.4}}^{+2.07}_{-1.84}$                     & ${2.29^{+0.93}_{-1.1}}^{+1.9}_{-2.0}$                 \\
                \hline
                {\boldmath$H_0$}                                   & ${67.88\pm0.39}^{+0.79}_{-0.77}$                 & ${67.96\pm0.39}^{+0.73}_{-0.76}$                 & ${67.98^{+0.37}_{-0.42}}^{+0.84}_{-0.77}$             \\
                {\boldmath$\Omega{}_{m}$}         & ${0.3082\pm0.0052}^{+0.010}_{-0.010}$            & ${0.3072\pm0.0051}^{+0.010}_{-0.0095}$           & ${0.3069\pm0.0053}^{+0.010}_{-0.011}$                 \\
                {\boldmath$\sigma_8$}             & ${0.757^{+0.053}_{-0.019}}^{+0.062}_{-0.092}$    & ${0.775^{+0.022}_{-0.016}}^{+0.034}_{-0.038}$    & ${0.767\pm0.016}^{+0.031}_{-0.032}$                   \\
                {\boldmath$S_8$}            & ${0.768^{+0.053}_{-0.022}}^{+0.066}_{-0.095}$    & ${0.785^{+0.022}_{-0.018}}^{+0.036}_{-0.042}$    & ${0.775\pm0.016}^{+0.030}_{-0.031}$                   \\ \hline 
                {\boldmath$\chi^2_{\rm min}$}           &
                 $ 4212.18 $  &  $4225.98$ 
                & $4226.74$
                \\
            \end{tabular}
        \end{ruledtabular}}
    \caption{The mean values and 1, 2$\sigma$ of the $\Lambda w$DM model for CMB+DESI+PP, CMB+DESI+PP+RSD and CMB+DESI+PP+RSD+WL datasets.}
    \label{tab:LwDM_DESI}
\end{table*}

\begin{table*}[!t]
    \centering
    \renewcommand\arraystretch{1.5}
    \setlength{\tabcolsep}{1mm}{
        \begin{ruledtabular}
            \begin{tabular}{cccc}
                Parameters                                                       & CMB+DESI+PP                                 & CMB+DESI+PP+RSD                                         & CMB+DESI+PP+RSD+WL                          \\
                \hline
                {\boldmath$100\omega{}_{b}$}    & ${2.246\pm0.012}^{+0.027}_{-0.025}$         & ${2.248\pm0.014}^{+0.027}_{-0.027}$                     & ${2.252\pm0.013}^{+0.027}_{-0.025}$         \\
                {\boldmath$\omega{}_{\mathrm{cdm}}$}      & ${0.11877\pm0.00082}^{+0.0016}_{-0.0016}$   & ${0.11848^{+0.00087}_{-0.00078}}^{+0.0016}_{-0.0017}$   & ${0.11793\pm0.00080}^{+0.0016}_{-0.0016}$   \\
                {\boldmath$100\theta{}_{s}$}      & ${1.04203\pm0.00027}^{+0.00053}_{-0.00055}$ & ${1.04202^{+0.00030}_{-0.00027}}^{+0.00056}_{-0.00062}$ & ${1.04207\pm0.00028}^{+0.00055}_{-0.00054}$ \\
                {\boldmath$\ln(10^{10}A_{s})$}              & ${3.054\pm0.014}^{+0.029}_{-0.028}$         & ${3.049\pm0.015}^{+0.030}_{-0.029}$                     & ${3.044\pm0.014}^{+0.029}_{-0.027}$         \\
                {\boldmath$n_{s}$}        & ${0.9685\pm0.0037}^{+0.0073}_{-0.0072}$     & ${0.9688\pm0.0035}^{+0.0068}_{-0.0065}$                 & ${0.9701\pm0.0035}^{+0.0069}_{-0.0066}$     \\
                {\boldmath$\tau{}_{\mathrm{reio}}$}          & ${0.0598\pm0.0073}^{+0.014}_{-0.014}$       & ${0.0580^{+0.0069}_{-0.0077}}^{+0.016}_{-0.015}$        & ${0.0564\pm0.0071}^{+0.014}_{-0.014}$       \\
                \hline
                {\boldmath$H_0$}                            & ${67.96\pm0.38}^{+0.76}_{-0.71}$            & ${68.08\pm0.38}^{+0.79}_{-0.71}$                        & ${68.33\pm0.37}^{+0.68}_{-0.71}$            \\
                {\boldmath$\Omega{}_{m}$}                & ${0.3072\pm0.0049}^{+0.0095}_{-0.0099}$     & ${0.3056\pm0.0049}^{+0.0095}_{-0.010}$                  & ${0.3023\pm0.0047}^{+0.0093}_{-0.0091}$     \\
                {\boldmath$\sigma_8$}                     & ${0.8115\pm0.0058}^{+0.012}_{-0.011}$       & ${0.8085\pm0.0060}^{+0.012}_{-0.011}$                   & ${0.8051\pm0.0056}^{+0.011}_{-0.011}$       \\
                {\boldmath$S_8$}                          & ${0.8212\pm0.0094}^{+0.019}_{-0.019}$       & ${0.8160^{+0.0099}_{-0.0090}}^{+0.018}_{-0.019}$        & ${0.8081\pm0.0090}^{+0.018}_{-0.017}$       \\     \hline 
                {\boldmath$\chi^2_{\rm min}$}           &
                 $ 4210.94 $  &  $4226.76$ 
                & $4232.58$
                \\
            \end{tabular}
        \end{ruledtabular}}
    \caption{The mean values and 1, 2$\sigma$ of the $\Lambda$CDM model for CMB+DESI+PP, CMB+DESI+PP+RSD and CMB+DESI+PP+RSD+WL datasets.}
    \label{tab:LCDM_DESI}
\end{table*}

\begin{table}[!t]
	\centering
	\renewcommand\arraystretch{1.2}
	\setlength{\tabcolsep}{3mm}{
		\begin{ruledtabular}
			\begin{tabular}{ccc}	
	Parameter &  $\Lambda w$DM & $\Lambda$CDM\\
	\hline
	{\boldmath$10^{2}\omega{}_{b }$} & $2.234\pm 0.036^{+0.069}_{-0.070}  $
	& $2.233\pm 0.036^{+0.070}_{-0.072}$ \\
	
	{\boldmath$\omega{}_{\rm cdm }$} & ${0.105^{+0.044}_{-0.025}}^{+0.046}_{-0.055}    $  &${0.101^{+0.046}_{-0.025}}^{+0.049}_{-0.053} $\\
	
	{\boldmath$H_0            $} & ${69^{+10}_{-8}}^{+20}_{-20}         $&${68^{+10}_{-9}}^{+20}_{-20}$\\
	
	{\boldmath$\sigma_8       $} & $0.822\pm 0.048^{+0.099}_{-0.091}            $&${0.809^{+0.043}_{-0.050}}^{+0.096}_{-0.090}$\\
	
	{\boldmath$10^{7}w_{\mathrm{dm} }$} & $< 5.02 < 9.51               $&  $-$\\
	
	{\boldmath$\Omega{}_{m }  $} & ${0.263^{+0.029}_{-0.036} }^{+0.066}_{-0.062}  $&${0.265^{+0.030}_{-0.037}}^{+0.069}_{-0.066} $\\
	
	{\boldmath$S_8            $} & $0.765\pm 0.021 ^{+0.042}_{-0.039}         $ & $0.755\pm 0.019^{+0.039}_{-0.037} $ \\ \hline
	{\boldmath$\chi^2_{\rm dm}$}  & $6.12$ & $6.18$
	 \end{tabular}
\end{ruledtabular}}
\caption{The mean values and 1, 2$\sigma$ of the $\Lambda$CDM model for RSD+WL datasets.}
\label{tab:RSD+WL}
\end{table}

\begin{table*}[!t]
	\renewcommand\arraystretch{1.1}
	\setlength{\tabcolsep}{0.6mm}{
		{\begin{tabular}{c c c }
				\hline
				\hline
Model &  Datasets & $ \triangle {\rm AIC}_{\Lambda w {\rm DM},\Lambda {\rm CDM}} $
\\ 
\hline

$\Lambda w$DM   &CMB  &   $2.60$
\\
$\Lambda w$DM&   CMB+SDSS+PP   &  $2.16$
\\
$\Lambda w$DM&  CMB+SDSS+PP+RSD  &   $-1.16$
\\
$\Lambda w$DM&  CMB+SDSS+PP+RSD+WL  &  $-6.40$
\\
$\Lambda w$DM&   CMB+DESI+PP   &  $3.24$
\\
$\Lambda w$DM&  CMB+DESI+PP+RSD  &  $1.22$
\\
$\Lambda w$DM&  CMB+DESI+PP+RSD+WL  &  $-3.84$
\\
$\Lambda w$DM&  RSD+WL  &  $1.94$
\\
\hline
\hline
\end{tabular}
\caption{ Difference of AIC values of the $\Lambda w$DM model with respect to the $\Lambda$CDM model for CMB, CMB+SDSS+PP, CMB+DESI+PP, CMB+SDSS+PP+RSD, CMB+DESI+PP+RSD, CMB+SDSS+PP+RSD+WL, and RSD+WL datasets.}
\label{tab:AIC}}}
\end{table*}

\section{concluding remarks}
The $\Lambda w$DM model, incorporating a cosmological constant and barotropic dark matter with a constant equation of state parameter $w_{\rm dm}$, demonstrates a potential pathway to alleviating the persistent $S_8$ tension between early- and late-universe cosmological probes. Our analysis, leveraging state-of-the-art datasets including Planck CMB, SDSS/DESI BAO, Pantheon+ supernovae, and KiDS-1000 weak lensing, reveals a marginally non-zero $w_{\rm dm}=2.7^{+2.0}_{-1.9}\times10^{-7}$(
at 95\% confidence level) for SDSS BAO data combined with all other non-BAO data, and  a marginally non-zero $w_{\rm dm}=2.29^{+1.9}_{-2.0}\times10^{-7}$(
at 95\% confidence level) for DES Y1 BAO data combined with all other non-BAO data.
In addition, we also find that, compared to $\Lambda$CDM, $\Lambda w$DM can relieve the $S_8$ tension from 
larger than $3\sigma$ to smaller than $1\sigma$. However, model comparisons based on AIC criteria indicate no decisive statistical preference for  $\Lambda w$DM over the standard model, with $\triangle$AIC values falling short of the threshold for robust evidence. Nevertheless, for CMB+SDSS+PP+RSD+WL datasets, the  $\Lambda w$DM model is close to being positively preferred over the  $\Lambda $CDM model, which is shown by the
absolute value of $\triangle$AIC that larger than 6. In the Future, we will address the critical limitation, i.e. the absence of applying the full WL likelihood for $\Lambda w$DM, by modifying halo model suitable for $\Lambda$CDM to fit the $\Lambda w$DM model. This will further test the viability of the $\Lambda w$DM model, offering deeper insights into whether dark matter is not cold.
\section*{Acknowledgments}
This work is supported by Guangdong Basic and Applied Basic Research Foundation (Grant No.2024A1515012573), the National key R\&D Program of China (Grant No.2020YFC2201600), National Natural Science Foundation of China (Grant No.12073088), and National SKA Program of China (Grant No. 2020SKA0110402).

\bibliographystyle{spphys}
\bibliography{LwDM}

\end{document}